\documentclass[twocolumn,prb]{revtex4-1}

\usepackage{amsmath}
\usepackage{amsfonts}
\usepackage{graphicx}
\usepackage{color}

\newcommand{\nearsites}[1]{\langle #1 \rangle}
\newcommand{\pdagger}{{\vphantom{\dagger}}}
\newcommand{\fref}[1]{fig.~\ref{#1}}

\newcommand{\eref}[1]{(\ref{#1})}
\newcommand{\ii}{\mathrm{i}\,}
\newcommand{\vect}[1]{\mathbf{#1}}
\newcommand{\const}{\mathrm{const}}

\newcommand{\chem}[1]{$\mathrm{#1}$}

\pdfoutput=1

\graphicspath{ {fig-png/}, {fig-eps/} }

\begin{document}
\title {Colossal magnetoresistance in topological Kondo insulator}
\author{Igor O. Slieptsov and Igor N. Karnaukhov}
\affiliation{G.V. Kurdyumov Institute for Metal Physics, 36 Vernadsky Boulevard, 03142 Kiev, Ukraine}
\begin{abstract}
Abnormal electronic properties of complex systems require new ideas concerning explanation of their behavior and possibility of realization. In this acticle we have shown that a colossal magnetoresistance is realized in the state of the topological Kondo insulator, that is similar to the Kondo insulator state in the Kondo lattice. The mechanism of the phenomenon is the following: in the spin gapless phase an external magnetic field induces the gap in the spectrum of spin excitations, the gap in the spectrum of fermions is opened due to a hybridization between spin and fermion subsystems at half filling, as the result the magnetic field leads to metal-insulator (or bad metal -- insulator) phase transition.
A model of the topological Kondo lattice defined on a honeycomb lattice is studied for the case when spinless fermion bands are half filled. It is shown that the hybridization between local moments and itinerant fermions should be understood as the hybridization between corresponding Majorana fermions of the spin and charge sectors. The system is a topological insulator, single fermion and spin excitations at low energies are massive. We will show that a spin gap induces a gap in the charge channel, it leads to an appearance of a topological insulator state with chiral gapless edge modes and the Chern number one or two depending on the exchange integrals' values. The~relevance of this to the~traditional Kondo insulator state is discussed.
\end{abstract}
\maketitle

\section{Introduction}

The problem of co-existence of itinerant and localized electrons represented via a single local magnetic moment
(the Kondo lattice problem) in crystals remains one of unsolved problems in the condensed matter physics.
A~state of Kondo insulator (KI) is realized in a Kondo lattice at the half filling.
It is defined by the Hamiltonian
${\cal H}_{\rm KL} = - t \sum_\sigma\sum_{\nearsites{i,j}}a^\dagger_{i\sigma} a^\pdagger_{j\sigma} +J\sum_j \vec{\sigma}_j \vec{S}_j$,
where the spin densities are coupled to the localized spins $\vec{S}_j$ through an antiferromagnetic exchange coupling $J$.
KIs~are heavy fermionic compounds, in which localized electrons hybridize with itinerant electrons and form a completely filled band of quasi-particles with spin and charge excitation gaps, localized electrons participate in the Fermi surface formation~\cite{ki-1,ki-2,ki-3,ki-4,ki-5}. The~numerical results~\cite{ki-1,ki-2,ki-5} show that the larger spin gap induces the gap in the charge channel and leads to an insulator state for odd number of itinerant electrons per local moment.
KI state is a result of strong interactions between itinerant and localized states of electrons. The~long-standing problems of whether localized electrons determine the volume of the Fermi surface and how the spin and charge gaps are formed in metal phase of itinerant electrons are unsolved.

Topological insulators (TI) have attracted a great attention from a view-point of a possible realization of topological phases in real compounds.
Topological insulators exhibit exotic physics, the research is motivated by both theoretical and experimental considerations~\cite{ex1-1,ex1-2,ex1-3}.
Traditionally, the studies of topological insulators have been made within the band theory of noninteracting fermions~\cite{km-1,km-2,km-3}.
According to~ref.~\cite{1,2,3,4}, a topological Kondo insulator (TKI) is an insulating state in heavy fermion
systems which appears as a result of a strong interaction between itinerant and localized electrons with a strong spin-orbit coupling. The~most salient example of the Kondo insulator is \chem{SmB_6}~\cite{ex-1,ex-2,ex-3,ex-4}.

The aims of the paper is to consider an academic problem of the Kondo insulator in the aspect of a realization of colossal magnetic resistance in new compounds.
In~contrast to the theory of the colossal magnetic resistance in manganites~\cite{man}, we consider the realization of the colossal magnetic resistance as the result of a nontrivial topology of the complex system.

\begin{figure}[b]
  \centering{\leavevmode}
  \includegraphics[width=0.7\linewidth]{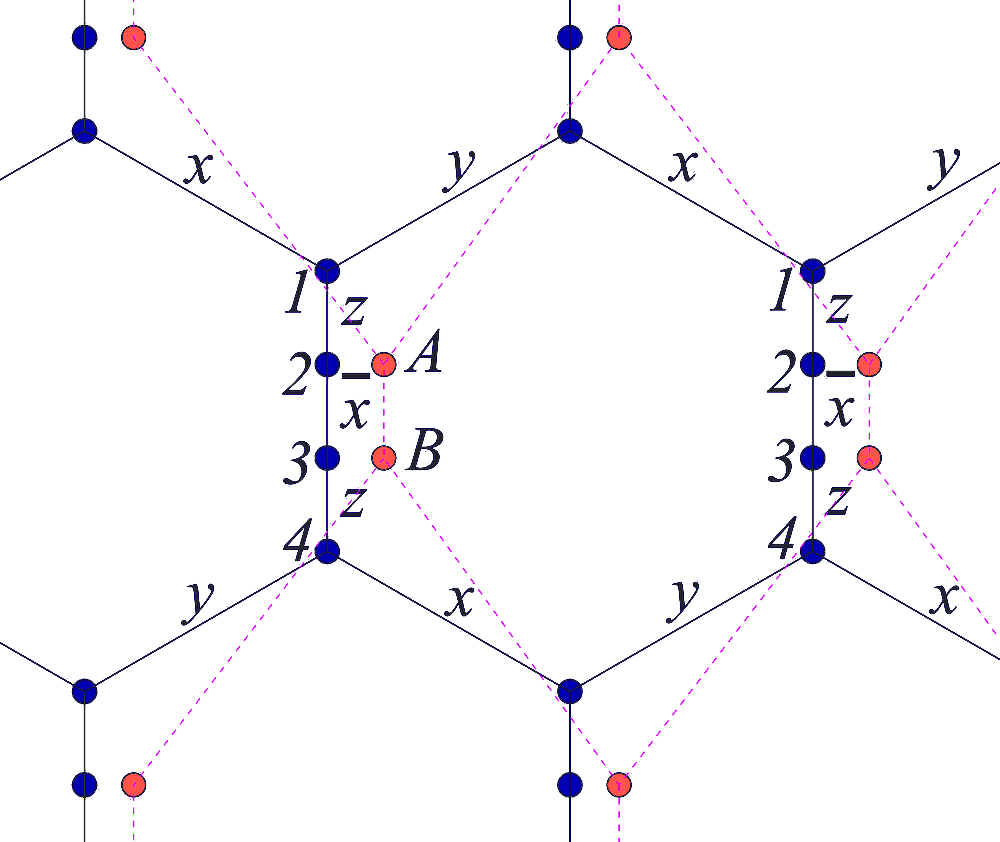}
  \caption{(Color online)
   Topological Kondo lattice, red sites $A$ and $B$ of fermion subsystem, blue sites with localized \mbox{spin-$\frac12$}, links of the types $x$, $y$, $z$ and $\bar{x}$
  }
  \label{fig:Model}
\end{figure}

We use a two-dimensional topological Kondo lattice (TKL) model based on recent publication~\cite{2I} in order to study a TKI state and show that the TKI state is similar to KI state in the Kondo lattice.
We~study the model of the TKL at the half filling and show that
a hybridization of itinerant fermions with local moments leads to an insulating state.
In~contrast to the state of traditional KI, the phase state of TKI is topologically nontrivial with different values of the Chern number~$C$.
A~weak magnetic field breaks the time reversal (TR) symmetry of the total system and induces the phase transitions to topological states.
Our aim is to demonstrate TKI state forming in the framework of the TKL model proposed.

\section{The topological Kondo lattice model}

\subsection{Fermion subsystem}

Let us start with a subsystem of spinless fermions, that at half-filling is a metal (i.\,e. a band structure has no gap), speaking more precisely, it is a bad metal, but not an insulator. The Hamiltonian of the fermion subsystem is defined on a honeycomb lattice~\cite{2I} shown in \fref{fig:Model} with two sites $A$ and $B$ per unit cell. We introduce both standard kinetic hopping and pairing terms for spinless fermions on nearest-neighbor lattice's sites with equal hopping and pairing amplitudes
\begin{equation}
  {\cal H}_{f} =- \ii t \sum_{\nearsites{i,j}}(a^\dagger_{i} a^\pdagger_{j} + a_{i} a_{j}) + h.c.,
  \label{eq:Hf}
\end{equation}
where $a_{j}^\dagger$ and  $a^\pdagger_{j}$ denote the creation and annihilation operators of spinless fermions at site $j$,
$t$ is a real-value nearest-neighbor hopping between two ordered neighbor sites $\nearsites{i,j}$ of the different types $A$ and~$B$.
Let us consider the presented term in~\eref{eq:Hf} denoting a hopping from $A$ to~$B$.

\begin{figure}[tb]
  \centering{\leavevmode}
  \includegraphics[width=0.8\linewidth]{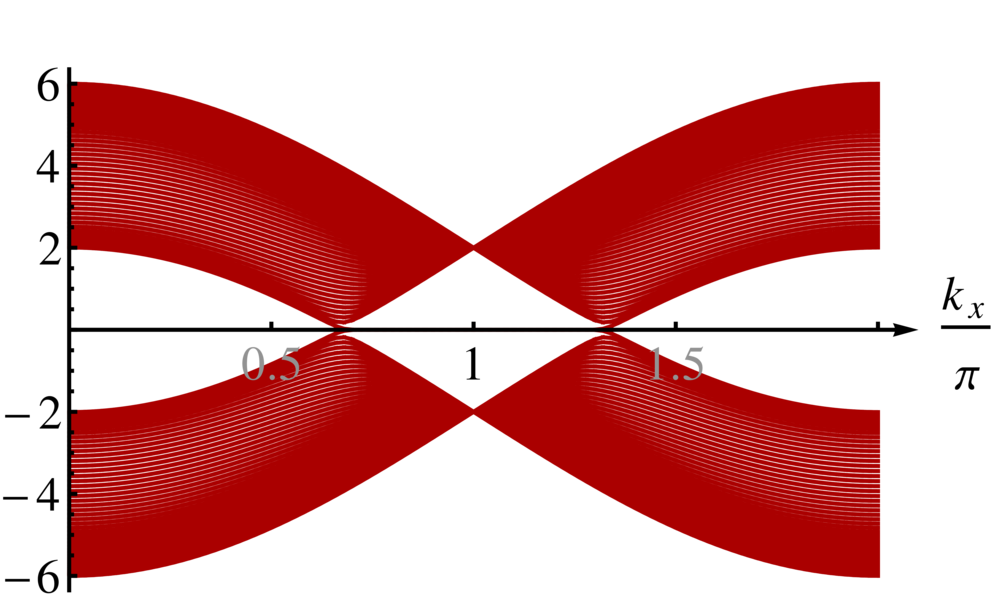}
  \caption{Energy levels of the fermion subsystem ${\cal H}_f$}
  \label{fig:FermionSubsystem}
\end{figure}

We introduce two types of the Majorana fermions on each site of the fermion sublattice using the relations
$$d_{j} = a^{\phantom{\dagger}}_{j} + a^\dagger_{j}\qquad\textrm{ and }\qquad g_{j}= \frac{a^{\phantom{\dagger}}_{j}-a^\dagger_{j}}{\ii}$$
and rewrite the Hamiltonian~\eref{eq:Hf} as ${{\cal H}_{f} =- \ii t \sum_{\nearsites{i,j}}d_{i} d_{j}}$.
It is reduced to a quadratic form of the Fermi operators
${\cal H}_{f}=-\ii \sum_{\textbf{k},\alpha,\beta}\varepsilon_{\alpha\beta} (\textbf{k})d^\dagger_{\textbf{k}\alpha} d^\pdagger_{\textbf{k}\beta}$
with the spectrum of one-particle excitations given by eigenvalues $\ii \varepsilon_{\alpha\beta} (\textbf{k})$, where
$d_{\textbf{k}\alpha}^\dagger=\frac{1}{\sqrt{2N}}\sum_sd_{s\alpha}\exp(\ii\vect{k} \vect{r}_s)$  and
$d_{\textbf{k}\alpha}=\frac{1}{\sqrt{2N}}\sum_sd_{s\alpha}\exp(-\ii\vect{k} \vect{r}_s)$
are the Fermi operators defined on the lattice with doubled unit cell~\cite{Kitaev}, $\textbf{k}$ is a~wave vector, $N$ is a~total number of unit cells, $s$ is an index of a cell, $\alpha,\beta\in\{A,B\}$.
Thus, in spite of the pairing terms in~\eref{eq:Hf}, the Fermi level is well defined and the energy levels of the Hamiltonian~\eref{eq:Hf} are the same as for the one-particle Hamiltonian with solely hopping terms. In the absence of an interaction, the fermion subsystem is always in non-topological phase, its gapless spectrum is shown in~\fref{fig:FermionSubsystem}.

\subsection{Spin subsystem}

We consider a variation of the Kitaev honeycomb model~\cite{Kitaev} for the spin subsystem: a model with di\-rec\-tion-depen\-dent exchange interactions with an additional $\bar{x}$-type interaction on bonds in $z$-direction. Explicitly, it is given by the Hamiltonian~\cite{2I}
\begin{multline}
    {\cal H}_{s} = \Delta_x \sum_{\nearsites{i,j}x} \sigma_i^x \sigma_j^x + \Delta_y \sum_{\nearsites{i,j}y} \sigma_i^y \sigma_j^y + \\
    \Delta_z \sum_{\nearsites{i,j}z} \sigma_i^z \sigma_j^z + I \sum_{\nearsites{i,j}{\bar{x}}} \sigma_i^x \sigma_j^x,
    \label{eq:Hs}
\end{multline}
consisting of exchange interaction $\Delta_\gamma\sigma_i^\gamma \sigma_j^\gamma$ with the exchange integrals $\Delta_\gamma$ between nearest sites $\nearsites{i,j}$ connected by $\gamma$-links, $\gamma=x,y,z$.
$\sigma_{j}^\gamma$~are the three Pauli operators at site~$j$.
Two additional sites (denoted by 2 and 3 in \fref{fig:Model}) are added to the unit cell, the exchange interaction between spins located at these sites is of the type~$x$ ($\sigma_i^x\sigma_j^x$) with the exhange integral~$I$.
For convenience it is supposed that all the exchange integrals are positive.

\begin{figure}[tb]
  \centering{\leavevmode}
  \includegraphics[width=0.48\linewidth]{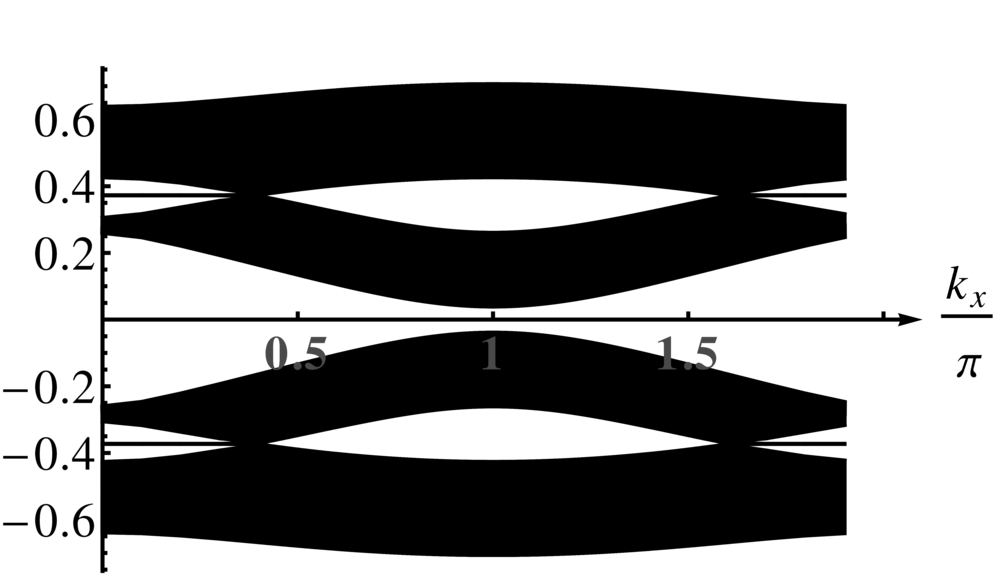}
  \includegraphics[width=0.48\linewidth]{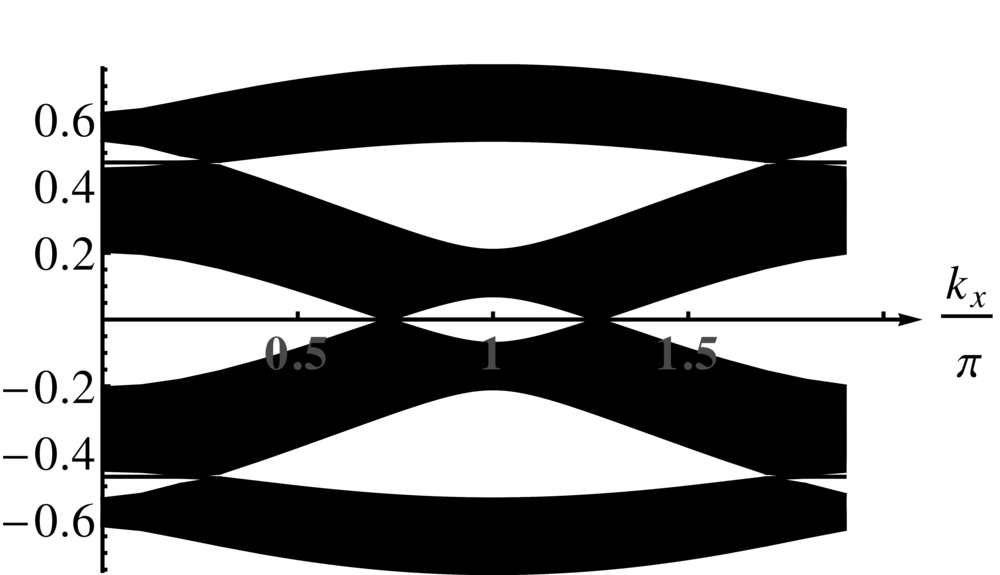}
  \caption{
    Energy bands of the localized spin-$\frac12$ subsystem in the absence of an external magnetic field, $\Delta_x=\Delta_y=t/4$, $\Delta_z = t/3$, a~gapped phase at~$I=t/6$ (left) and a~gapless phase at~$I=t/3$ (right)
  }
  \label{fig:SpinSubsystem}
\end{figure}

The model is solved analytically for the uniform configurations, due to the translational invariance of the lattice.
The vortex-free configuration is the ground state of the model~\cite{Kitaev,2I}.
The ground-state phase diagram of the spin subsystem coincides with the one of the Kitaev model in coordinates $I \Delta_x$, $I \Delta_y$, $ \Delta_z^2$, it is drawn on the plane $I\Delta_x + I\Delta_y +\Delta_z^2 = \const$~\cite{2I}. Topologically trivial (gapped, with~$C_s=0$) and nontrivial (gapless, with~$C_s=1$) phases are separated with lines of topological phase transitions.
The topologically nontrivial phase state is stable due to TR symmetry, the energy spectrum of Majorana fermions can not open gaps without breaking TR symmetry. A weak magnetic field ${\cal H}_H = -\vec{H}\sum_{j}\vec{\sigma}_j$ breaks TR symmetry of the spin subsystem and opens the gap in the spectrum of Majorana fermions. It can be taken into account using perturbation theory, the gap in the spectrum is defined by a scale $h\sim H^3/\Delta^2$ ($\Delta_\gamma \sim \Delta$). The structure of the spectra of spin excitations is plotted for each phase in \fref{fig:SpinSubsystem} for comparison.

\subsection{Interaction}

We consider an adiabatic connection of the spin and fermion subsystems taking into account an interaction term in the total Hamiltonian of the TKL ${\cal H}={\cal H}_{f}+{\cal H}_{s}+{\cal H}_{i}$. Let us define $ {\cal H}_i$ according to ref.~\cite{2I}
\begin{equation}
    {\cal H}_{i} = \lambda \sum_{\alpha} (2 a^\dagger_{\alpha}a^\pdagger_{\alpha}-1)\sigma^y_{\alpha},
    \label{eq:Hint}
\end{equation}
where the coupling parameter $\lambda$ defines a contact interaction of spinless fermions with local moments on sites $\alpha$ ranging all contacting pairs $(A,2)$ and $(B,3)$, see \fref{fig:Model}.

The evolution of the subsystems from an initial state (the noninteracting fermion and spin subsystems which have been considered above) to a final TKI state defined by the Hamiltonian ${\cal H}$ will be considered below. For convenience, an energy scale is set to be $t$ and henceforth all the exchange integrals and the coupling constant $\lambda$ are measured in unit of~$t$.

The Hamiltonian ${\cal H}$ can be exactly diagonalized using the representation of the Pauli operators in terms of a related set of the Majorana fermions $b_j^\gamma$ and $c_j$ with the substitution $\sigma^\gamma_j=\ii b^\gamma_jc^{\phantom{\gamma}}_j$~\cite{Kitaev}
\begin{multline}
  {\cal H} = -\frac{\ii}{2} \sum_{\beta=x,y,z,\bar{x}}\sum_{\nearsites{i,j}\beta} A_{ij}^\beta c_i c_j \\
    - \ii t \sum_{\nearsites{i,j}} d_i d_j + \lambda\sum_{\alpha} g_\alpha d_\alpha c_\alpha b_\alpha^y,
\label{eq:H}
\end{multline}
where  $A_{ij}^\gamma = \Delta_\gamma u_{ij}^\gamma$ for the directed links $\gamma=x,y,z$ and $A_{ij}^{\bar x}=Iu_{ij}^{\bar x}$ for the intercalated $\bar{x}$-link, $u_{ij}^\gamma = -u_{ji}^\gamma = \ii b_{i}^\gamma b_{j}^\gamma$.

The operators $u_{ij}^{\beta}$ and plaquette operators (similar to the Kitaev model~\cite{Kitaev}) are constants of motion with eigenvalues $\pm 1$. Other operators $u_\alpha=\ii g_{\alpha}b_{\alpha}^y$ describe contact pairing of spin and fermion Majorana fermions on sites $\alpha$, they are the constants of motion with eigenvalues $\pm 1$. The equation~\eref{eq:H} with conserved $u_{ij}^{\beta}$ and $u_\alpha$ is the Hamiltonian of free Majorana fermions in the background frozen ${\mathbb{Z}}_2$ gauge fields. Thus, the charge and spin Majorana fermions hybridize and open gaps in charge and spin channels.

\section{Topological Kondo insulator state}

When we considered the forming of the TKI state in the introduction, we drew the analogy to the KI state realized on the Kondo lattice.
The KI state is realized on the Kondo lattice at half-filling of electrons, this phase state is a spin-charge insulator with larger spin gap.
At the half-filling in the absence of an interaction between subsystems the spin subsystem~\eref{eq:Hf} is a bad metal (\fref{fig:FermionSubsystem}), when spin subsystem~\eref{eq:Hs} exhibits in topological trivial (gapped spectrum, ${C_s=0}$) and nontrivial (gapless spectrum, ${C_s=1}$) states (\fref{fig:SpinSubsystem}).

The contact interaction~\eref{eq:Hint} breaks both particle-hole and TR symmetries of the model Hamiltonian and gives a Kitaev-like ground-state phase diagram~\cite{Kitaev} of the system. The hybridized spectrum of spin-charge excitations defines a large volume of the Fermi sea.

Let us consider an adiabatic connection of the subsystems via the interaction~\eref{eq:Hint} and an evolution of the ground state of the system along $\lambda$ directions with the spin Hamiltonian's parameters fixed. Note that the values of the exchange integrals define only initial and final states of the system.
We will consider the model with a $xy$-symmetric spin-exchange interaction $\Delta_x = \Delta_y = \Delta \neq \Delta_z$ without loss of generality for different values of~$I$. When $|I|\leq \Delta_z^2 / (2 |\Delta|)$, the non\-interacting spin subsystem has a gapless spectrum of the Majorana fermions, and it is gapped for other values of $I$. We shall use $\Delta=t/4$ and $\Delta_z=t/3$ to demonstrate general results.

\begin{figure}[tb]
  \centering{\leavevmode}
  \includegraphics[width=0.48\linewidth]{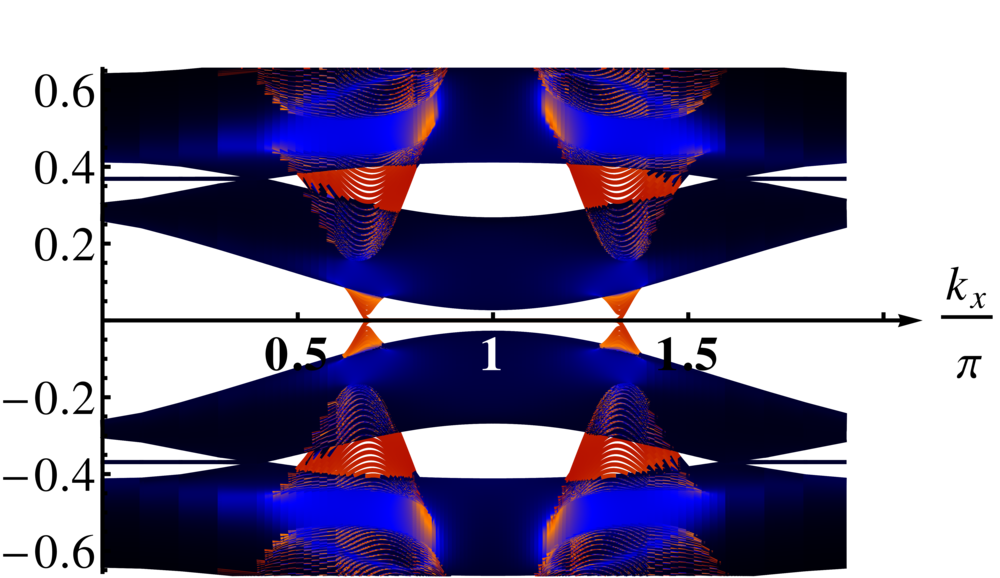}
  \includegraphics[width=0.48\linewidth]{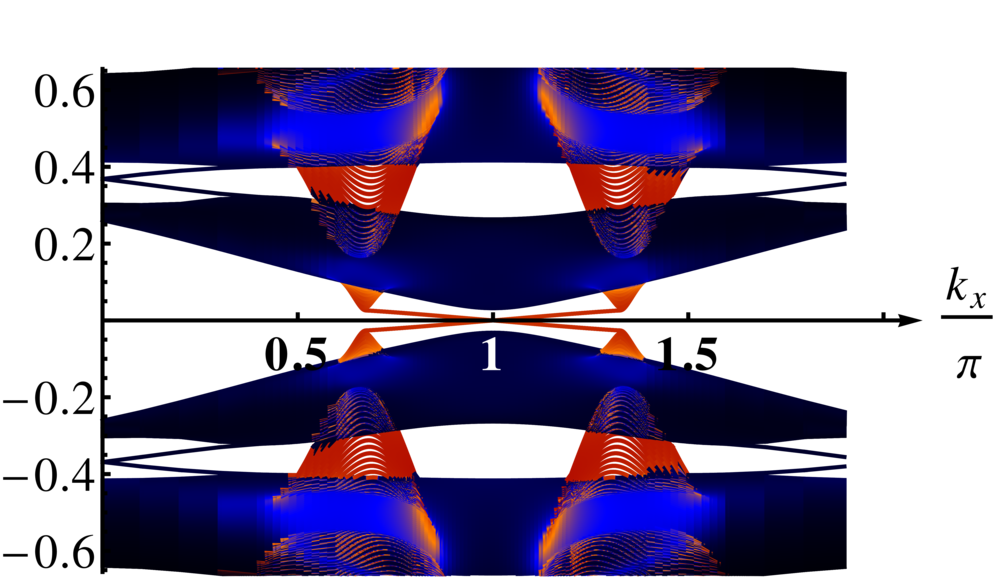}
  \caption{(Color online)
    Energy spectrum calculated in the case of topological trivial state of noninteracting spin subsystem at $\Delta_x=\Delta_y=t/4$, $\Delta_z=t/3$, $I=t/6$, $\lambda=0.1t$ with the wave vector along the zig-zag boundary,
    $h=0$ (left) and $h=0.1t$ (right)
  }
  \label{fig:GappedPhase}
\end{figure}

\begin{figure}[tb]
  \centering{\leavevmode}
  \includegraphics[width=0.48\linewidth]{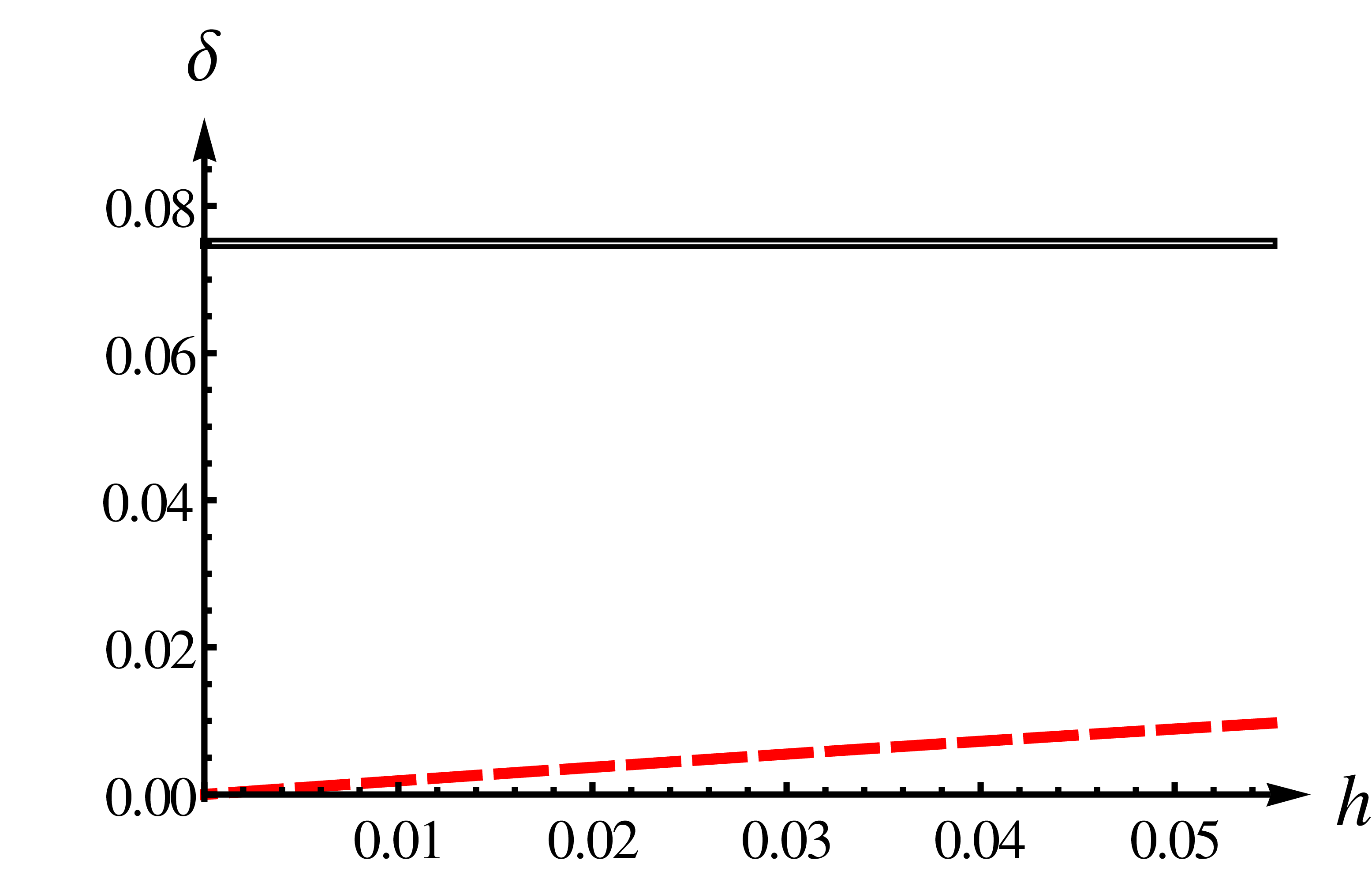}
  \includegraphics[width=0.48\linewidth]{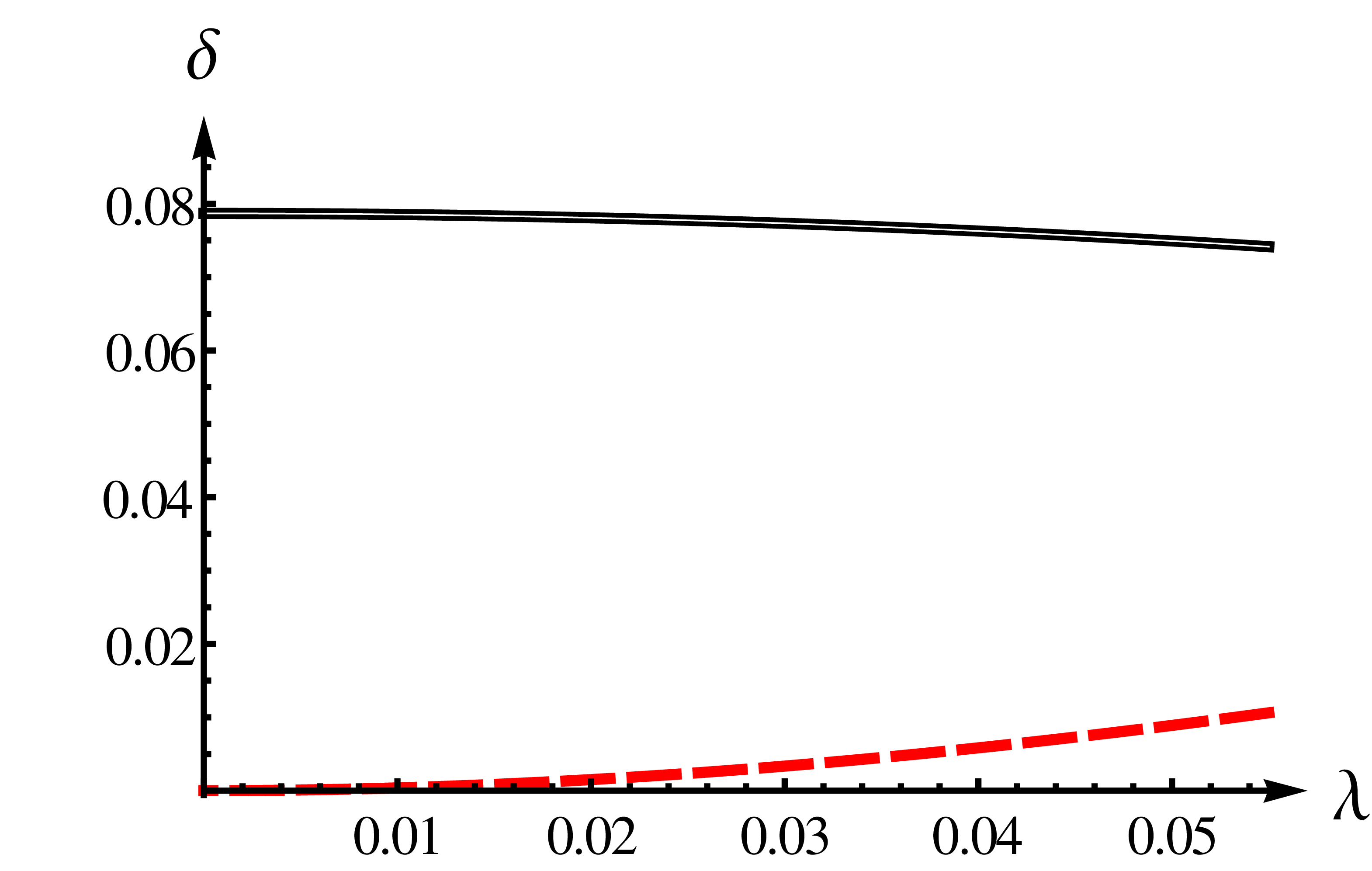}
  \caption{(Color online) The spin (black solid line) and charge (dashed red line) gaps as functions of $\lambda$ for $h=0.05t$ (right) and of $h$ for $\lambda=0.05t$ (left) calcutated at $\Delta_x=\Delta_y=t/4$, $\Delta_z=t/3$, $I=t/6$.
  }
  \label{fig:GappedPhase1}
\end{figure}

\begin{figure*}[tb]
  \centering{\leavevmode}
  \includegraphics[width=0.3\linewidth]{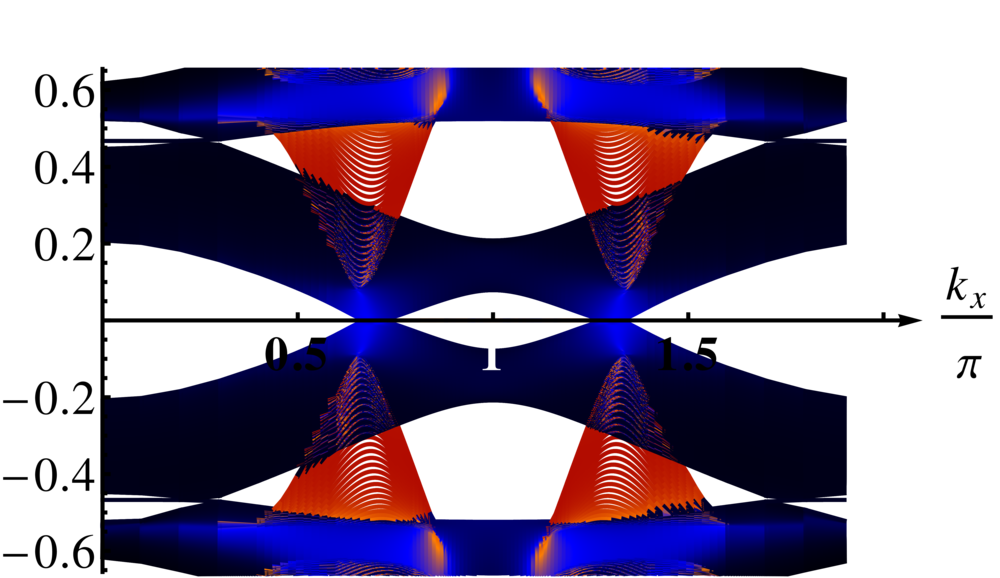}
  \includegraphics[width=0.58\linewidth]{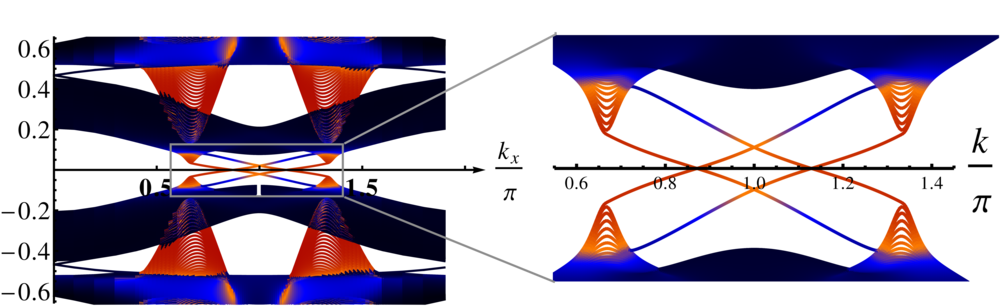}
  \caption{(Color online) Energy spectrum calculated in the case of topological nontrivial state of noninteracting spin subsystem at $\Delta_x=\Delta_y=t/4$, $\Delta_z=t/3$, $I=t/3$, $\lambda=0.1t$ with the wave vector along the zig-zag boundary;
    $h=0$~(left),
    $h=0.1t$~(right)
  }
  \label{fig:GaplessPhase}
\end{figure*}

Let us start with the topologically trivial phase state of noninteracting spin subsystem with $C_s=0$.
When $h=0$ and $\lambda \neq 0$, the spin gap does not open a charge gap in the charge channel of excitations, the phase state of fermion subsystem is not changed, see \fref{fig:GappedPhase}(left). A weak magnetic field ($h \neq 0$) opens the charge gap and stabilizes the TKI phase, see \fref{fig:GappedPhase}(right). The larger spin gap weakly depends on the magnitude of the magnetic field and the coupling constant, whereas the low-energy fermion gap is proportional to $h$ and $\lambda^2$ (see \fref{fig:GappedPhase1}).
Calculations of the Chern number and edge states show that  $C = 1$ and two chiral gapless edge modes are realized in the TKI spectrum. The edge modes have a fermionic origin, but one can not say about pure charge fermion states
when spin and fermionic excitations are hybridized.
These edge modes verify the obtained result $C=1$, the state is topological. Such behavior of the system is characteristic for arbitrary values of the exchange integrals with $2 |\Delta| |I|\geq \Delta_z^2$ corresponding to the gapped phase of the spin Hamiltonian~\eref{eq:Hs}.

\begin{figure}[tb]
  \centering{\leavevmode}
  \includegraphics[width=0.48\linewidth]{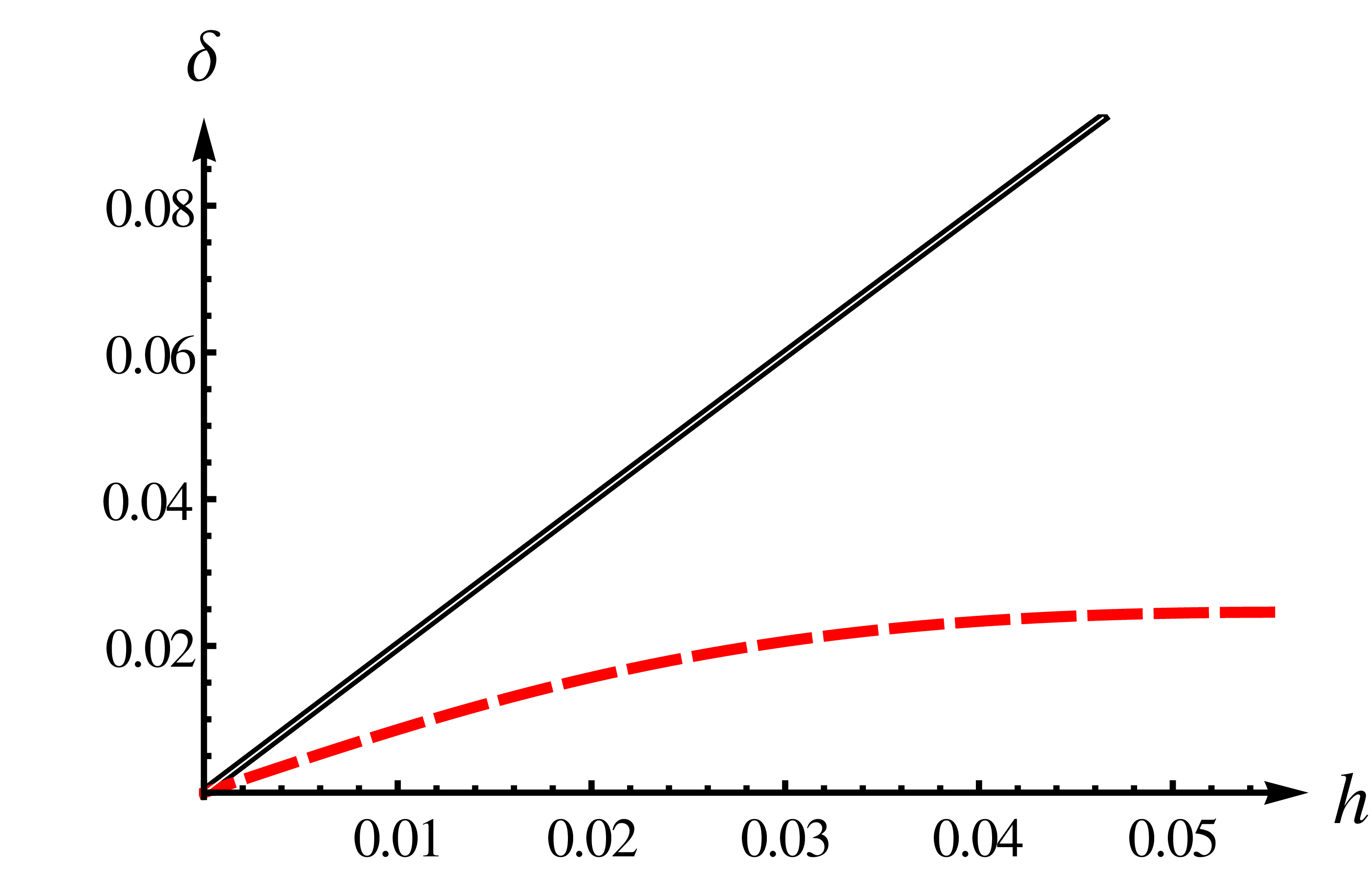}
  \includegraphics[width=0.48\linewidth]{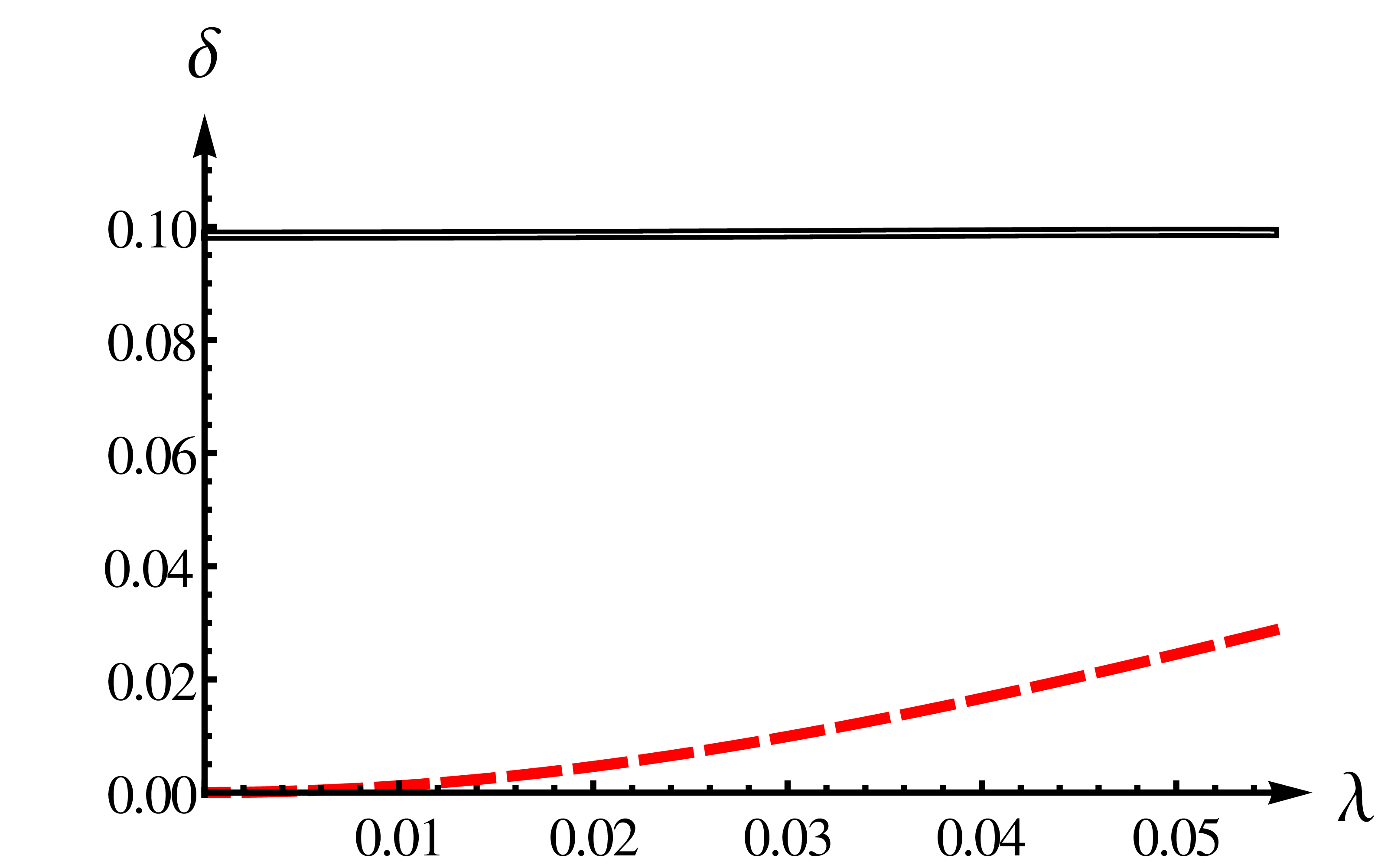}
  \caption{(Color online) The spin (black solid line) and charge (dashed red line) gaps as functions of $ \lambda$ for $h=0.05t$ (right) and $h$ for $\lambda=0.05t$ (left)
  calculated at $\Delta_x=\Delta_y=t/4$, $\Delta_z=t/3$, $I=t/3$.
  }
  \label{fig:GaplessPhase1}
\end{figure}

Now consider the exchange integrals' values of the topologically nontrivial phase of \eref{eq:Hs} with $C_s=1$.
The gap in the spectrum of the spin subsystem is induced by a weak magnetic field $h\neq 0$
that stabilizes a spin-topological nontrivial phase (see \fref{fig:GaplessPhase}).
The gap in the spectrum of spin excitations $\sim h$ (see \fref{fig:GaplessPhase1}, right) forms a gap in the spectrum of charge excitations $\sim (h \lambda^2)/(\Delta t)$ (see \fref{fig:GaplessPhase1}, left). The~TKI state is topologically nontrivial with $C=2$, the edge modes are associated with both spin and charge subsystems (see \fref{fig:GaplessPhase}). The gapless edge modes have a hybrid structure, they connect spin and fermion subbands in different regions of the spectrum.

The point of the topological phase transition from bad-metal state to TKI state has coordinates $\lambda=0$ and $h=0$, therefore a resistivity of TKL strongly depends on the applied magnetic field $\vec{H}$. In the weak magnetic field a gap is opened, the system exhibits the topological phase transition from bad metal to insulator state. The gap in the charge channel increases with the magnitude $H$ of the magnetic field. In the region of weak magnetic field the TKI state offers a giant magnetoresistivity. The~hybridization of itinerant spinless fermions and local moments form the larger volume of the Fermi sea, local moments contribute to the volume of Fermi sea an additional unit $\int_{\rm BZ} d\vect{k} = n + 1$, where $n$ is a density of fermions and integral is taken over the Brillouin zone of the system.

A take-home definition of topological Kondo insulator can be said in the following way.
TKI is a topological spin-charge insulator, wherein a spin gap in the spin-charge hybridized spectrum induces a gap in the charge channel. Gapless spin-charge chiral edge modes with the nontrivial Chern number ${C=1}$ or ${C=2}$ characterize the topological state of~TKI.
This state exhibits a giant magnetoresistivity in the region of weak magnetic field.

\section{Conclusions}

We have considered a formation of TKI state in the honeycomb TKL model which ground state is a spin-charge insulator at the half-filling and shown a mechanism of the hybridization between spin and fermion excitations.
The scenario of the TKI state forming that is realized in the proposed model, it is analogues to the scenario of the KI state.
The~first, the Majorana fermions that describe local moments hybridize with the Majorana fermions of itinerant spinless fermions contributing to the volume of Fermi sea.
The~second, the charge gap is opened due to the spin gap induced by a magnetic field which breaks TR symmetry of the spin subsystem.
Additionally, the TKI state is topologically nontrivial with the Chern number ${C=1}$ or ${C=2}$ depending on the values of the exchange integrals.
The~gapless edge modes form a surface subband of chiral Majorana fermions.

The main result of the paper is the following: topolological complex systems, such as the topological Kondo lattice, can possess abnormal properties, such as colossal magnetoresistivity. A giant magnetoresistivity in weak magnetic fields is a dramatic peculiarity of the topological phase transition in the TKI state.


\begin{thebibliography}{31}
\bibitem{ki-1}
    G. Aeppli and Z. Fisk, Comm. Condens. Matter Phys. \textbf{16}, 155 (1992).
\bibitem{ki-2}
    H. Tsunetsugu, M. Sigrist and K. Ueda, Rev. Mod. Phys. \textbf{69}, 809 (1997).
\bibitem{ki-3}
    I.N. Karnaukhov, Phys. Rev. B, \textbf{56}, 4313(R) (1997).
\bibitem{ki-4}
    I.N. Karnaukhov, Phys. Rev. B, \textbf{57}, 3863 (1998).
\bibitem{ki-5}
    P. Riseborough, Adv. Phys. \textbf{49}, 257 (2000).
\bibitem{ex1-1}
    D. Hsieh, D. Qian, L. Wray, Y. Xia, Y. S. Hor, R. J. Cava and M. Z. Hasan, Nature \textbf{452}, 970 (2008).
\bibitem{ex1-2}
    Y. Xia, D. Qian, D. Hsieh, L. Wray, A. Pal, H. Lin, A. Bansil, D. Grauer, Y. S. Hor, R. J. Cava and M. Z. Hasan, Nat. Phys. \textbf{5}, 398 (2009).
\bibitem{ex1-3}
    Y. L. Chen, J. G. Analytis, J.-H. Chu, Z. K. Liu, S.-K. Mo, X. L. Qi, H. J. Zhang, D. H. Lu, X. Dai, Z. Fang, S. C. Zhang, I. R. Fisher, Z. Hussain and Z.-H. Shen, Science \textbf{325}, 178 (2009).
\bibitem{km-1}
    Haldane F. D. M., Phys. Rev. Lett., 61 (1988) 2015.
\bibitem{km-2}
    C. L. Kane and E. J. Mele, Phys. Rev. Lett. \textbf{95}, 146802 (2005).
\bibitem{km-3}
    I.N. Karnaukhov and I.O. Slieptsov, Eur.Phys.J. B, \textbf{87}, 230 (2014).
\bibitem{1} Alexandrov V., Dzero M. and Coleman P., Phys.Rev. Lett., 111 (2013) 226403.
\bibitem{2} Dzero M., Sun K., Galitski V. and Coleman P., Phys. Rev. Lett., 104 (2010) 106408.
\bibitem{3} Dzero M., Sun K., Coleman P. and Galitski V., Phys. Rev. B, 85 (2012) 045130.
\bibitem{4} Lu F., Zhao J.-Z., Weng H.-M., Fang Z. and Dai X., Phys. Rev. Lett., 110 (2013) 096401.
\bibitem{ex-1}
    Kim D. I., Thomas S., Grant T., Botimer I., Fisk Z. and Iing Xia, Scientific Reports, 3 (2013) 3150.
\bibitem{ex-2}
    Wolgast S., Kurdak C., Sun K., Allen I. W., Kim D. I. and Fisk Z., Phys. Rev. B, 88 (2013) 180405(R).
\bibitem{ex-3}
    Zhang X, Butch N. P., Syers P., Ziemak S., Greene R. L. and Paglione I. P., Phys. Rev. X, 3 (2013) 011011.
\bibitem{ex-4}
    Ciomaga Hatnean M., Lees M. R., Paul D. McK. and Balakrishnan G., Scientific Reports, 3 (2013) 3071.   
    \bibitem{man} Cepas O., Krishnamurthy H.R. and Ramakrishnan T.V.,
     Phys.Rev.Lett. 94,  247207 (2005).
\bibitem{2I} I.N. Karnaukhov and I.O. Slieptsov, Europhys.Lett.  \textbf{109}, 57005 (2015).
\bibitem{Kitaev} A.Yu. Kitaev, Ann.Phys.  321, 2 (2006).
\end{thebibliography}
\end{document}